# Circular Photocurrents in Centrosymmetric Semiconductors with Hidden Spin Polarization

Kexin Wang[1], Butian Zhang[1, *], Chengyu Yan[1], Luojun Du[2, 3], Shun Wang[1, †]

[1]*MOE Key Laboratory of Fundamental Physical Quantities Measurement and Hubei Key Laboratory of Gravitation and Quantum Physics, PGMF and School of Physics, Huazhong University of Science and Technology, Wuhan 430074, China*

[2]*Beijing National Laboratory for Condensed Matter Physics; Key Laboratory for Nanoscale Physics and Devices, Institute of Physics, Chinese Academy of Sciences, Beijing, 100190, China*

[3]*School of Physical Sciences, University of Chinese Academy of Sciences, Beijing 100190, China*

## Abstract

Centrosymmetric materials with site inversion asymmetries possess hidden spin polarization, which remains challenging to be converted into spin currents because the global inversion symmetry is still conserved. This study demonstrates the spin-polarized DC circular photocurrents (CPC) in centrosymmetric transition metal dichalcogenides (TMDCs) at normal incidence without applying electric bias. The global inversion symmetry is broken by using a spatially-varying circularly polarized light beam, which could generate spin gradient owing to the hidden spin polarization. The dependences of the CPC on electrode configuration, illumination position, and beam spot size indicate an emergence of circulating electric current under spatially inhomogeneous light, which is associated with the deflection of spin-polarized current through the inverse spin Hall effect (ISHE). The CPC is subsequently utilized to probe the spin polarization and ISHE under different excitation wavelengths and temperatures. The results of this study demonstrate the feasibility of using centrosymmetric materials with hidden spin polarization and non-vanishing Berry curvature for spintronic device applications.

**Introduction**

Spin and valley, as intrinsic degrees of freedom of electrons, are being investigated as new information carriers for next-generation devices[1,2]. Fundamental challenges in spintronics involve the injection, manipulation, and measurements of spin/valley polarization. A feasible method is to employ the circular photocurrent (CPC), i.e., to apply the circularly polarized light for spin/valley injection and measure the helicity-dependent photocurrent. CPC has been demonstrated in quantum wells[3], Weyl semimetals[4–7], and semiconducting transition metal dichalcogenides (TMDCs)[8–10]. The amplitudes of the such spin/valley-related photocurrents can be utilized to probe spin polarization[11,12], fermion chirality[4], Berry curvatures[6], etc.

Despite emerging in a variety of materials, a common view is that spin- or valley-related CPC of nonmagnetic materials embodies a breaking of spatial inversion symmetry[13]. An intuitive approach is to use intrinsic non-centrosymmetric crystals or structures, such as $T_d$-phase TMDCs[6,14–16], monolayer 2H-phase TMDCs[9,10], and surfaces of topological insulators[17]. As for centrosymmetric crystals, researchers develop a series of approaches to break the spatial inversion symmetry, including interlayer twisting, heterostructure construction, strain engineering, and the application of external electric field[18]. A typical example is 2H-phase TMDCs. Although monolayer and thin 2H-TMDCs with odd number of layers lacks inversion symmetry, the even layers, thick multilayers, and bulk crystals are considered to be centrosymmetric. As expected, the spin or valley-related CPC component in bilayer and thick multilayer TMDCs were non-vanishing only when the structural inversion symmetry is broken by external electric field through ionic gating[19–21] or constructing Schottky barriers[22,23].

A decade ago, it was established that hidden spin polarization could exist in centrosymmetric crystals with atomic site inversion asymmetry, including but not limited to bulk silicon and bulk 2H-phase TMDCs[24]. Although bulk 2H-TMDCs, belonging to $D_{6h}$ point group, have zero net spin polarization, opposite spin polarizations arising from local Dresselhaus effect are spatially localized in individually non-centrosymmetric $α$ and $β$ sectors with $D_{3h}$ point group symmetry. Subsequently, the hidden spin polarization has been directly observed in 2H-phase $WSe_2$, $MoS_2$, and $MoTe_2$ by spin- and angle-resolved photoemission spectroscopy (spin-ARPES)[25–28] and is used to explain the helicity-dependent photoluminescence[29,30] and elliptically polarized terahertz emission[31].

The existence of spin/valley polarization is necessary but not sufficient for spintronic devices. Although the helicity-dependent photoluminescence shows the possibility for photoexcitation of the spin-polarized carriers in bilayer or multilayer 2H-TMDCs, it was found that the intrinsic spin or valley-related CPC current and Kerr rotation angle vanish without symmetry breaking[19,22,32]. This problem limits the spintronic applications of multilayer 2H-TMDCs despite their higher conductivity and stability compared to their monolayer counterparts. The same challenge exists for other centrosymmetric structures with hidden spin polarizations.

In this study, intrinsic spin-polarized DC CPC is demonstrated in centrosymmetric

TMDCs including thick multilayer 2H-phase MoTe$_2$, MoS$_2$ and WSe$_2$ at normal incidence under no external electric bias. Instead of breaking the structural inversion symmetry of the material, we introduce the symmetry breaking of the system by adopting a spatially-varying beam profile. The manifestation of hidden spin-polarization by CPC is attributed to the inverse spin Hall effect (ISHE), which converts the light-induced spin gradient to charge current. The induced circulating current is confirmed by the dependence of CPC on the electrode configuration and illumination position. Subsequently, the CPC is utilized to reveal the information on spin polarization of 2H-TMDCs by evaluating its wavelength-dependence and temperature-dependence. The results demonstrate feasibility of using multilayer 2H-TMDCs for practical spintronic applications and contributes to the ongoing exploration of materials for spintronic applications.

## Results and Discussion

The crystal structure of 2H-TMDCs is depicted in Fig. 1a. The upper and lower layers of the unit cell are respectively represented as $\alpha$ sector and $\beta$ sector, which are inversion symmetric about the red point in the middle. $\alpha$ and $\beta$ sectors are also in a 180° rotation relative to each other and thus show reversed K and K' valleys. The valence band of each layer splits into VB1 and VB2 due to the Ising spin-orbit coupling (SOC). This energy splitting allows one to choose an excitation wavelength for selective excitation of carriers from VB1. Assuming that the $\alpha$ sector and $\beta$ sector were two independent monolayers, the right circular polarized σ+ light can excite the spin-down states ($|\downarrow\rangle$) at K valleys of both the $\alpha$ sector ($K_\alpha$) and $\beta$ sector ($K_\beta$) but no excited states from K'$_\alpha$ and K'$_\beta$ valley, as depicted in Fig. 1b. No spin-up ($|\uparrow\rangle$) states are excited and the spin polarization ratio is 100%. For an actual multilayer 2H-TMDC crystal, the spin polarization ratio of K point is 28%-91% revealed by spin-ARPES[25–28,33]. As schematically shown in Fig. 1c, interlayer coupling between $\alpha$ and $\beta$ sectors causes additional splitting of VB1 and VB2 and also the mixing of $|\uparrow\rangle$ and $|\downarrow\rangle$ states at K and K' valleys[27,29] (see details in Supplementary Fig. S1). In the picture of hidden spin polarization, carriers from $\alpha$ and $\beta$ sectors have opposite spin polarizations at either K or K' valley. Moreover, carriers from the $K_\alpha$, $K_\beta$, K'$_\alpha$, and K'$_\beta$ valleys inherit their original optical selection rules under σ+/σ- optical excitation[29]. Consequently, σ+ light with a specific wavelength could excite predominantly $|\downarrow\rangle$ states and a smaller portion of $|\uparrow\rangle$ states at K ($K_\alpha$) and K' ($K_\beta$) valleys of VB1, as indicated by the unfaded arrows. Correspondingly, σ- optical excitation is expected excite carriers at K'$_\alpha$ and K'$_\beta$ valleys and generate opposite spin polarization.

To obtain spin polarized photocurrent, we propose a method to convert the differences in spin of carriers under σ+/σ- optical excitation into differences in electric currents. Specifically, a focused Gaussian laser beam is normally incident on centrosymmetric TMDCs to induces a gradient distribution of spin-polarized charge carriers proportional to the local light intensity, leading to a radially diffused spin current (light blue arrows in Fig. 1e). The spin current could be converted to a transverse charge current if ISHE is applicable in 2H-TMDCs, as we will discuss later. Driven by the ISHE, the moving carriers from the spin current will deviate from their original paths and acquires a velocity component along the circumference (black arrows in Fig. 1e), forming a circulating charge current. Using a pair of head-to-head electrodes deviating from center of a circle will collect uncompensated charge current with its sign determined by the charge circulating direction. In this way, the angular momentum of photons can be converted into the circular motion of electrons.

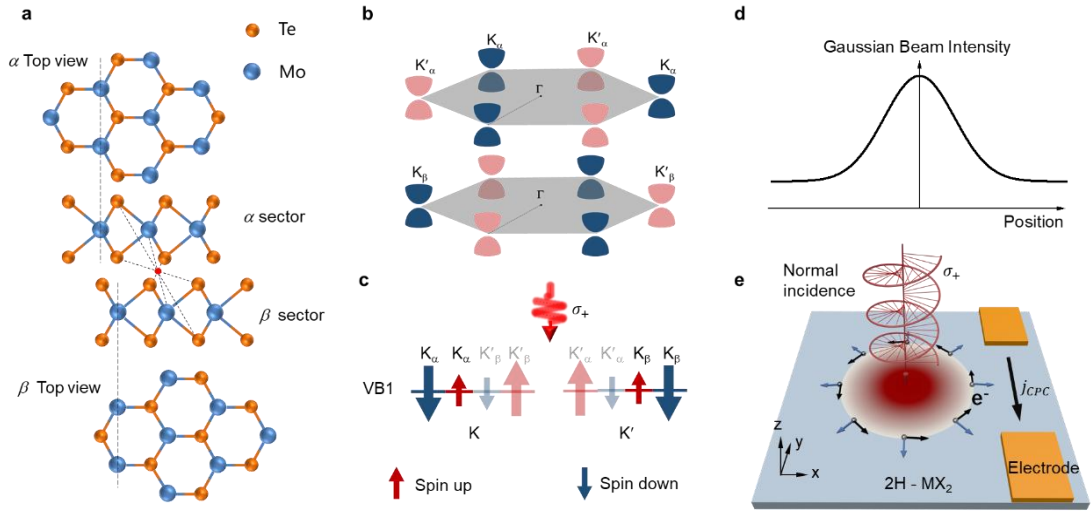

**Fig. 1 Proposed mechanism of spin-polarized CPC in multilayer 2H-MoTe$_2$. a,** Crystal structures of 2H-MoTe$_2$. **b,** Brillouin zone and band structure of TMDC without interlayer coupling. Right circular polarized light σ+ excites spin down states at K$_α$ valleys of α sector and K$_β$ valleys of β sector. **c,** Hybridization of spin states in valence band VB1 of multilayer TMDC with interlayer coupling. σ+ excites a majority of spin-down electrons and a minority of spin-up electrons, leading to a net spin polarization. **d,** The spatial distribution of laser intensity for Gaussian beams. **e,** Schematic diagram depicting the creation of a spin current induced by the intensity gradient of the Gaussian beam, subsequently converting into the circulating charge current. The light blue arrows indicate the direction of pure spin current, while the black arrows represent the direction of charge current.

Based on the above mechanism, an experimental set-up is constructed for measuring the CPC of centrosymmetric 2H-TMDCs, as schematically shown in Fig. 2a. The multilayer 2H-phase MoTe$_2$ (>20 nm) was adopted as a representative centrosymmetric semiconductor due to a high hidden spin polarization ratio[33] and large SOC splitting[34]. The 2H phase is confirmed by XRD, Raman spectroscopy and electrical characteristics of the MoTe$_2$ transistors (Supplementary Fig. S2-S4). A circular Gaussian laser spot is normally incident on MoTe$_2$ through an objective lens. A band-pass filter of 1100±10 nm is applied for the selectively exciting the A exciton at K and K′ point and the light polarization is modulated by rotating a quarter-wave plate.

All of the measurements are carried out at room temperature under vacuum conditions. Two types of devices made of multilayer MoTe$_2$ with different electrode configurations, namely A and B, are shown in Fig. 2b and 2c. Device A adopts two head-to-head electrodes contacting opposite edges of the channel for the observation of CPC. Device B, as a control, uses a pair of parallel electrodes across the width of the MoTe$_2$ channel. Spot **a** and spot **b** of two devices are respectively illuminated for obtaining polarization-dependent photocurrents, as shown in Fig. 2d and 2e. Upon illumination on spot **a**, the periodic fluctuations with a period of π correspond to a circular polarization dependence, while the differences between the photocurrents under σ- light (45°, 225°) and those under σ+ light (135°, 315°) indicate the emergence of CPC. Circular photon drag effect (CPDE) is suggested as a dominant source of CPC in monolayer TMDCs when the obliquely incident light is applied.[8] In this study, the possibility of CPDE-induced CPC is naturally ruled out at an incidence normal to the crystal plane[35,36]. Moreover, although weak spatial symmetry breaking can be extrinsically introduced into thin layers via spontaneous doping from substrate or external gating, this interface effect is estimated to be negligible in thick multilayers spanning tens of nanometers. Even for bilayer 2H-TMDC, a potential difference between the top and bottom layers is estimated to be only on the order of ~meV by using solid gating, which is too low to induce spin/valley Hall effect[37]. Notably, no external electric bias or gating voltages is applied and the possibility of phase transition in $MoTe_2$ induced by laser irradiation[38,39] has been eliminated (Supplementary Fig. S3). Therefore, the CPC arises from the intrinsic properties of 2H-$MoTe_2$. The phenomena observed in Fig. 2d is applicable to other multilayer 2H-TMDCs. Non-vanishing CPC is also observed in thick multilayer 2H-$MoS_2$ and 2H-$WSe_2$ (see Supplementary Fig. S5). When spot **b** of device B is illuminated, the photocurrent shows almost no CPC but only a linear polarization dependence, as shown in Fig. 2e. This result is consistent with previous studies on multilayer 2H-phase TMDC devices using an electrode configuration same with B, where CPC diminishes under normal incidence or without ionic gating[19–21]. As shown in Supplementary Fig. S6b, CPC at oblique incidence is non-vanishing for device B under the focused laser beam, consistent with the observation in bilayer $MoS_2$[40] with same electrode configuration. But the magnitude of CPC at oblique incidence decreases when the light intensity gradient is reduced by increasing the off-focus amount (see Supplementary Fig. S6c).

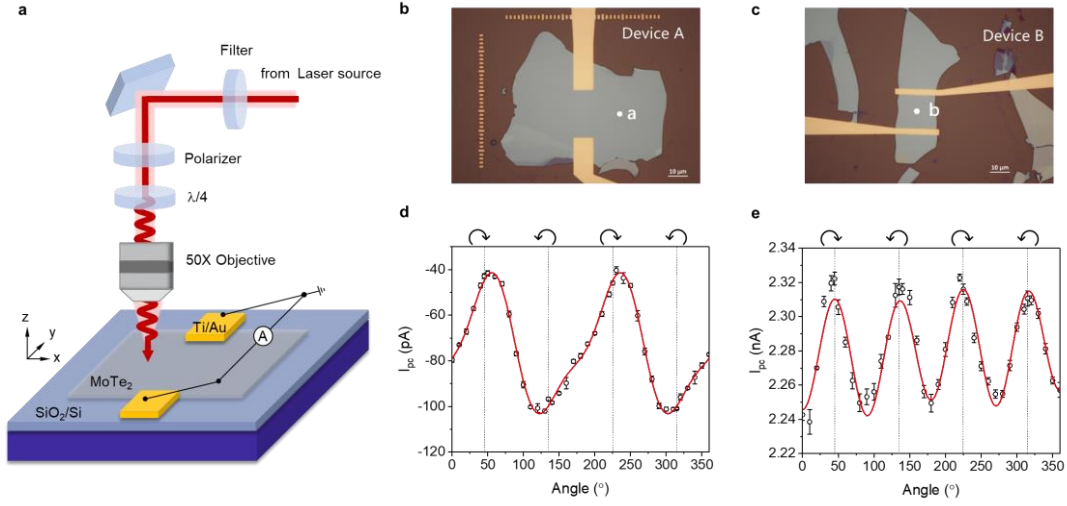

**Fig. 2 MoTe$_2$ devices and polarization-dependent photocurrent measurements. a,** Schematic experimental set-up for measuring helicity-dependent photocurrents. **b, c,** Optical photographs of multilayer devices with different electrode configurations. **e, f,** Photocurrents respectively measured from device in 2b/2c as rotation angle of the quarter wave plate is adjusted and spot **a** /**b** is illuminated. No bias or gate voltages is applied. Red curves were fitted by using Eq. (1).

The mechanism of CPC in device A is investigated by analyzing its dependence on illumination position and spot size. Fig. 3a demonstrates a series of polarization-dependent photocurrent ($I_{PC}$) curves when moving the laser spot along the vertical bisector line on device A (X position). The total photocurrent ($I_{PC}$) can be described as,

$$I_{PC}(\theta) = C \sin(2\theta) + L \sin(4\theta + \delta) + I_0 \qquad (1)$$

where $\theta$ is the rotation angle of quarter-wave plate, $C$ and $L$ account for helicity-dependent and linear polarization-dependent components, and $I_0$ is the polarization-independent background current. After fitting the $I_{PC}$ curves by using Eq. (1), the $C$ values are illustrated as a function of X position in Fig. 3b. Along X, the $C$ value shows a positive peak and a negative peak on opposite sides of the midpoint. Near the midpoint or away from the electrode pair, the $C$ value diminishes. Same results have been observed for a series of 2H-MoTe$_2$ devices made from 20-65 nm flakes. This distinctive spatial pattern of C component indicated a circulating charge current round the laser spot, which has been recognized as the characteristic of the ISHE. Previously, the spatially dispersive CPC have been observed from quantum well[41,42], topological insulator Bi$_2$Se$_3$[43], InN[44], T$_d$-MoTe$_2$[45], and ReS$_2$[46]. Herein, the emergence of ISHE in multilayer 2H-MoTe$_2$ indicate the existence of non-vanishing Berry curvature, which could arise from SOC or hidden Berry curvatures[47]. In TMDC systems with weak Rashba SOC, the spin Hall effect was attributed to Ising SOC[48], which is also suggested as a possible origin of the ISHE observed in this study. The out-of-plane spin polarization is orthogonal to both the radial spin current and the local electric current, which fulfills the requirement for spin-to-charge conversion in crystals with more than one mirror planes of symmetry[49,50]. On the other hand, hidden Berry curvature has recently been observed in bulk 2H-WSe$_2$ by ARPES[51], which may theoretically cause a Hall effect due to the same sign of Berry-curvatures for valleys selected by circularly polarized light [47,52]. However, further study is required to determine the origin of the non-vanishing Berry curvatures.

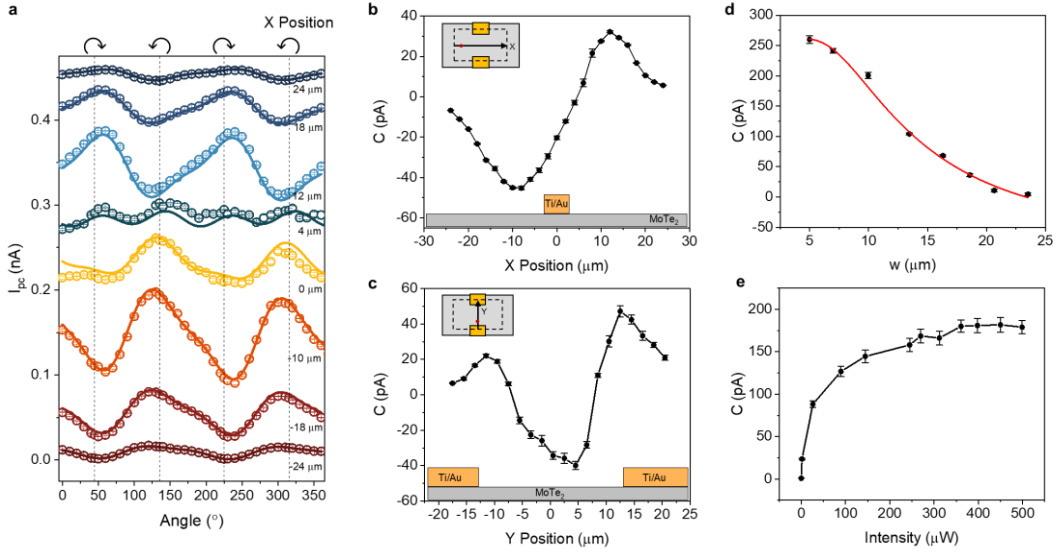

**Fig. 3 CPC dependent on illumination position, Gaussian spot size, and laser intensity.**

**a,** Photocurrent for different illumination location as a function of the quarter-wave plate angle. The circular dots represent experimental data, while the solid line represents the curve fitted using Eq. (1). **b,** Fitted C as a function of X position along the perpendicular bisector of channel. **c,** Fitted C as a function of Y position along the channel. The insets in (b) and (c) depict the relative displacement of laser spot with respect to the device. The yellow and gray areas represent the metal electrode and the $MoTe_2$ sample, respectively. **d,** The spot size dependence of fitted C under 1100 nm illumination at a fixed position. The $w$ is standard deviation of Gaussian distribution correlated with the spot size. The red curve is guideline for eyes. **e,** The power dependence of fitted C at minimum spot size under 1100 nm illumination. The measurements were carried out without applying bias or gate voltage.

An interesting comparison to our results is from a previous study on $MoTe_2$ of other crystal phases. Under the excitation of focused Gaussian beam profiles, a same condition with our study, Zhurun Ji et al.[16] observed no CPC in 1T'-$MoTe_2$ and circulating CPC in $T_d$-$MoTe_2$ and $Mo_{0.9}W_{0.1}Te_2$. They attribute the CPC to circular photogalvanic effect (CPGE) arising from antisymmetric terms in the nonlinear conductivity tensor under broken inversion symmetry, while the terms vanish in the inversion symmetric 1T'-MoTe$_2$. For 2H-MoTe$_2$ used in this study, the CPC is unlikely to originate from the antisymmetric conductivity terms because the CPGE tensor for 2H-TMDCs is zero[53]. Different from 1T'-MoTe$_2$ exhibiting inversion symmetry for both global crystal and each single layer, multilayer 2H-MoTe$_2$ crystals is centrosymmetric but shows site inversion asymmetry, which allows the generation of spin-polarized CPC.

The CPC observed along laser scan of the channel (Y position) exhibits a peak at both ends near the electrode, shown in Fig. 3c, which is due to the establishment of a Schottky barrier at the sample-electrode interface. The generation of CPC due to the existence of Schottky barrier has been observed in MoSe$_2$[23], Si nanowire[54], and semimetal Cd$_3$As$_2$[55], which is considered a third-order nonlinear effect[15]. The photocurrent profile is not perfectly symmetrical, showing a non-zero photocurrent at the midpoint during scan along X and Y directions. This could potentially be due to irregularities of the sample and the imperfect spot shape of the laser beam.

The effect of light intensity gradient is demonstrated by examining how Gaussian spot size influences the CPC. As illustrated in Fig. 3d, the enlargement of the spot size with fixed illumination intensity leads to a gradual decline in CPC until its eventual dissipation (Supplementary Fig. S7). The $w$ is standard deviation of Gaussian distribution correlated with the spot size. This observation indicated the pivotal role of the intensity gradient of the Gaussian light spot. When increasing the light intensity, as shown in Fig. 3e, $C$ firstly increase and then reach a saturation. The saturation is attributed to the gradual enlargement of the absorption saturation region in the Gaussian spot center as light intensity increases, limiting the spin gradient.

To determine the energy band origin of the spin photocurrent, the spectral response of CPC was acquired. Photocurrents $I_{PC}$ as a function of the quarter-wave plate angle were recorded at a fixed illumination position and different the incident wavelengths ranging from 700 nm-1400 nm, as illustrated in Fig. 4a. Extracted from these photocurrent curves, the coefficient $C$ as a function of incident wavelengths is depicted in Fig. 4b. Either when left or right side of the channel is illuminated, the magnitude of $C$ attains its maximum value at a wavelength of 1100 nm, which is near resonance with A exciton at the K point of ~1.1 eV[56,57]. When the incident wavelength deviates from 1100 nm to a longer wavelength, the magnitude of $C$ wanes as a result of the attenuated absorbance. Shorter incident wavelengths also correspond to smaller magnitudes of $C$ which is attributed to a lowering of the spin polarization. As the center wavelength is tuned from 1100 nm to 700 nm by using a filter with FWHM of 20 nm, the total net spin contains differently weighted contributions from A exciton, B exciton, indirect bandgap transition at 0.52 Γ-K point, and direct bandgap transition at K point. Among them, B exciton contributes an opposite spin relative to that of A exciton, 0.52 Γ-K transition has zero spin polarization, bandgap transition at K point contributes a spin polarization same with that of A exciton at the absorption edge and zero spin at shorter wavelengths. This could explain the vanishing of $C$ at 800 nm and shorter wavelength. The wavelength dependence of $C$ indicates that A excitons emerge as pivotal players in the generation of CPC[8,19].

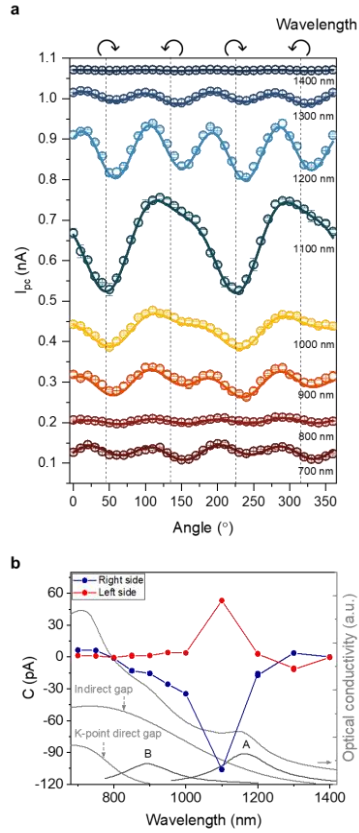

**Fig. 4 Wavelength dependence of the CPC. a,** Photocurrent as a function of the waveplate angle at different incident wavelengths from 700 nm to 1400 nm. **b,** The incident wavelength dependence of the fitted coefficient C. Red line and blue line represent C values when left-side spot and right-side spot is respectively illuminated, as labeled in the inset figure. Grey curves are optical conductivities of 10 nm-thick 2H-MoTe$_2$ adapted from Ref [56].

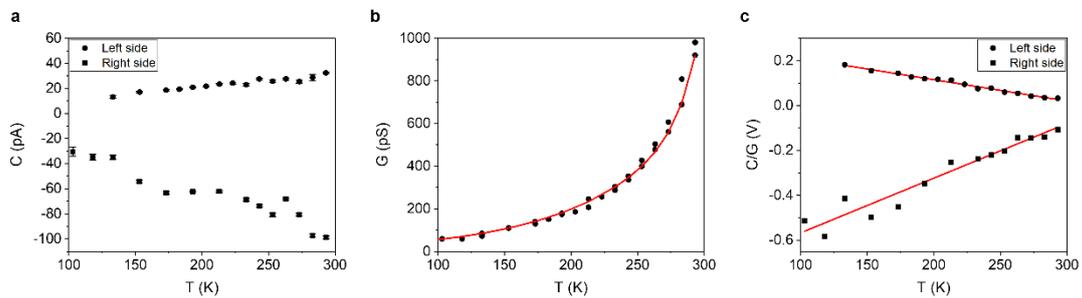

**Fig. 5 Temperature dependence of the CPC**. **a,** Fitted parameters C as a function of temperature. **b,** Electrical conductance G of device as a function of temperature. **c,** Temperature dependence of the C/G. Red curve and straight lines are guidelines for eyes.

Next, the temperature dependence of CPC is investigated. As depicted in Fig. 5a, the magnitudes of $C$ for left and right illumination spots diminish with a reduction in temperature. Different from 1T$_d$-MoTe$_2$ device, in which the CPC decreases with temperature due to the Lifshitz transition[14], a different mechanism is expected for 2H-phase MoTe$_2$. To examine the influences from mobility and contact resistance, the temperature dependence of the conductance $G$ derived from I-V curves (see Supplementary Fig. S8) was examined. As shown in Fig. 5b, $G$ declines more rapidly with decreasing temperature than the magnitude of $C$. To eliminate the influences of conductance, we extracted $C/G$ ratios to represent the electrical potential that drives the swirling current. As shown in Fig. 5c, the magnitude of $C/G$ rises when the temperature is decreased from 293 K to 103 K. The change in $C/G$ reflects the temperature effects on spin polarization ratio and spin or valley lifetime. In TMDCs, the spin polarization ratio reduces at lower temperatures due to the increase in interlayer coupling and almost unchanged SOC[27,29]. Given the positive correlation between temperature and $C/G$, we infer that the temperature effect on spin/valley scattering and relaxation prevails that on spin polarization ratio.

## Conclusion

In summary, spatially-dispersive CPC has been observed in multilayer 2H-MoTe$_2$ device without applying any external electric bias. In order to achieve the spin-polarized CPC, it was found necessary to use unevenly-distributed light beam to break the global inversion symmetry. The maximum amplitude of CPC is located at 1100 nm, suggesting the A excitons at K and K′ valley as the primary source of the photocurrent. Accordingly, the CPC is identified as intrinsic spin photocurrent originating from the hidden spin polarization and ISHE. Moreover, the observation of CPC under inhomogeneous light could be extended to multilayer 2H-MoS$_2$, 2H-WSe$_2$ and potentially other centrosymmetric materials with hidden spin polarization and non-vanishing Berry curvature, demonstrating the feasibility of using a new library of materials for valleytronics and spintronics devices.

## Methods

**Fabrication of MoTe$_2$ devices**

MoTe$_2$ nanoflakes were prepared by mechanical exfoliation from commercial bulk crystals (hq graphene) on 300 nm SiO$_2$/Si substrates. The electrodes were predefined by e-beam lithography and then deposited with 5/60 nm Ti/Au by electron beam evaporation.

**Characterizations of MoTe$_2$**

Raman spectra were obtained by using a lab-equipped Raman spectrometer consisting of a 488 nm laser, a 100× lens, a Princeton Instruments SP-2500 spectrometer and a Pixis 100 CCD camera. The thickness of MoTe$_2$, WSe$_2$, and MoS$_2$ was measured by an atomic force microscope (AFM, Bruker Dimension Edge).

**Photocurrent measurements**

The incident light from a supercontinuum laser (YSL photonics. SC-5) went through a single bandpass filter (Andover Corp. FS20-25), a linear polarizer, and a rotatable quarter-waveplate (Union Optic. WPA4420-900-2100, WPA4420-650-900) to modulate its polarization. The 1100 nm laser was focused by a 50× objective lens to an FWHM ~12 μm spot. The intensity distribution of the beam spot was measured using a CMOS beam profiler (WinCamD-LCM). The power of the left- and right- circularly-polarized light measured by a power meter showed a difference of less than 0.1%. Scanning photocurrent image was collected with a motorized two-dimensional displacement stage (S&I GmbH). Electrical measurements were carried out in vacuum at room temperature with an Keysight B1500A semiconductor parameter analyzer.

## Acknowledgements

We acknowledge the support from the National Natural Science Foundation of China (12074134).

## References

1. Mak, K. F., Xiao, D. & Shan, J. Light–valley interactions in 2D semiconductors. *Nat. Photonics* **12**, 451–460 (2018).

2. Li, L. *et al.* Room-temperature valleytronic transistor. *Nat. Nanotechnol.* **15**, 743–749 (2020).

3. Ganichev, S. D. & Prettl, W. Spin photocurrents in quantum wells. *J. Phys. Condens. Matter* **15**, R935–R983 (2003).

4. Ma, Q. *et al.* Direct optical detection of Weyl fermion chirality in a topological semimetal. *Nat. Phys.* **13**, 842–847 (2017).

5. Rees, D. *et al.* Helicity-dependent photocurrents in the chiral Weyl semimetal RhSi. *Sci. Adv.* **6**, eaba0509 (2020).

6. Xu, S.-Y. *et al.* Electrically switchable Berry curvature dipole in the monolayer topological insulator WTe$_2$. *Nat. Phys.* **14**, 900–906 (2018).

7. De Juan, F., Grushin, A. G., Morimoto, T. & Moore, J. E. Quantized circular photogalvanic effect in Weyl semimetals. *Nat. Commun.* **8**, 15995 (2017).

8. Quereda, J. *et al.* Symmetry regimes for circular photocurrents in monolayer MoSe$_2$. *Nat. Commun.* **9**, 3346 (2018).

9. Liu, L. *et al.* Electrical Control of Circular Photogalvanic Spin-Valley Photocurrent in a Monolayer Semiconductor. *ACS Appl. Mater. Interfaces* **11**, 3334–3341 (2019).

10. Eginligil, M. *et al.* Dichroic spin–valley photocurrent in monolayer molybdenum disulphide. *Nat. Commun.* **6**, 7636 (2015).

11. Yu, J. *et al.* Observation of current-induced spin polarization in the topological insulator Bi$_2$Te$_3$ via circularly polarized photoconductive differential current. *Phys. Rev. B* **104**, 045428 (2021).

12. Ma, Q., Krishna Kumar, R., Xu, S.-Y., Koppens, F. H. L. & Song, J. C. W. Photocurrent as a multiphysics diagnostic of quantum materials. *Nat. Rev. Phys.* **5**, 170–184 (2023).

13. Xu, H., Wang, H., Zhou, J. & Li, J. Pure spin photocurrent in non-centrosymmetric crystals: bulk spin photovoltaic effect. *Nat. Commun.* **12**, 4330 (2021).

14. Lim, S., Rajamathi, C. R., Süß, V., Felser, C. & Kapitulnik, A. Temperature-induced inversion of the spin-photogalvanic effect in WTe$_2$ and MoTe$_2$. *Phys. Rev. B* **98**, 121301 (2018).

15. Ma, J. *et al.* Circular photogalvanic effect from third-order nonlinear effect in 1T'-MoTe$_2$. *2D Mater.* **8**, 025016 (2021).

16. Ji, Z. *et al.* Spatially dispersive circular photogalvanic effect in a Weyl semimetal. *Nat. Mater.* **18**, 955–962 (2019).

17. McIver, J. W., Hsieh, D., Steinberg, H., Jarillo-Herrero, P. & Gedik, N. Control over topological insulator photocurrents with light polarization. *Nat. Nanotechnol.* **7**, 96–100 (2012).

18. Du, L. *et al.* Engineering symmetry breaking in 2D layered materials. *Nat. Rev. Phys.* **3**, 193–206 (2021).

19. Guan, H. *et al.* Photon wavelength dependent valley photocurrent in multilayer MoS$_2$. *Phys. Rev. B* **96**, 241304 (2017).

20. Yuan, H. *et al.* Generation and electric control of spin–valley-coupled circular photogalvanic current in WSe$_2$. *Nat. Nanotechnol.* **9**, 851–857 (2014).

21. Guan, H. *et al.* Inversion Symmetry Breaking Induced Valley Hall Effect in Multilayer WSe$_2$. *ACS Nano* **13**, 9325–9331 (2019).

22. Lee, J., Mak, K. F. & Shan, J. Electrical control of the valley Hall effect in bilayer MoS$_2$ transistors. *Nat. Nanotechnol.* **11**, 421–425 (2016).

23. Quereda, J. *et al.* The role of device asymmetries and Schottky barriers on the helicity-dependent photoresponse of 2D phototransistors. *Npj 2D Mater. Appl.* **5**, 13 (2021).

24. Zhang, X., Liu, Q., Luo, J.-W., Freeman, A. J. & Zunger, A. Hidden spin polarization in inversion-symmetric bulk crystals. *Nat. Phys.* **10**, 387–393 (2014).

25. Riley, J. M. *et al.* Direct observation of spin-polarized bulk bands in an inversion-symmetric semiconductor. *Nat. Phys.* **10**, 835–839 (2014).

26. Razzoli, E. *et al.* Selective Probing of Hidden Spin-Polarized States in Inversion-Symmetric Bulk $MoS_2$. *Phys. Rev. Lett.* **118**, 086402 (2017).

27. Zhang, Y. *et al.* The origin of the band-splitting and the spin polarization in bulk 2H-$WSe_2$. *Appl. Phys. Lett.* **122**, 142402 (2023).

28. Bussolotti, F., Maddumapatabandi, T. D. & Goh, K. E. J. Band structure and spin texture of 2D materials for valleytronics: insights from spin and angle-resolved photoemission spectroscopy. *Mater. Quantum Technol.* **3**, 032001 (2023).

29. Liu, Q., Zhang, X. & Zunger, A. Intrinsic Circular Polarization in Centrosymmetric Stacks of Transition-Metal Dichalcogenide Compounds. *Phys. Rev. Lett.* **114**, 087402 (2015).

30. Jones, A. M. *et al.* Spin–layer locking effects in optical orientation of exciton spin in bilayer $WSe_2$. *Nat. Phys.* **10**, 130–134 (2014).

31. Huang, Y. *et al.* Hidden spin polarization in the centrosymmetric $MoS_2$ crystal revealed via elliptically polarized terahertz emission. *Phys. Rev. B* **102**, 085205 (2020).

32. Mak, K. F., McGill, K. L., Park, J. & McEuen, P. L. The valley Hall effect in $MoS_2$ transistors. *Science* **344**, 1489–1492 (2014).

33. Tu, J. *et al.* Direct observation of hidden spin polarization in 2H-$MoTe_2$. *Phys. Rev. B* **101**, 035102 (2020).

34. Liu, G.-B., Shan, W.-Y., Yao, Y., Yao, W. & Xiao, D. Three-band tight-binding model for monolayers of group-VIB transition metal dichalcogenides. *Phys. Rev. B* **88**, 085433 (2013).

35. Jiang, C. *et al.* Helicity-dependent photocurrents in graphene layers excited by midinfrared radiation of a $CO_2$ laser. *Phys. Rev. B* **84**, 125429 (2011).

36. Shalygin, V. A., Moldavskaya, M. D., Danilov, S. N., Farbshtein, I. I. & Golub, L. E. Circular

photon drag effect in bulk tellurium. *Phys. Rev. B* **93**, 045207 (2016).

37. Wu, Z. *et al.* Intrinsic valley Hall transport in atomically thin $MoS_2$. *Nat. Commun.* **10**, 611 (2019).

38. Fukuda, T. *et al.* Photoinduced Structural Dynamics of $2H-MoTe_2$ Under Extremely High-Density Excitation Conditions. *J. Phys. Chem. C* **127**, 13149–13156 (2023).

39. Cho, S. *et al.* Phase patterning for ohmic homojunction contact in $MoTe_2$. *Science* **349**, 625–628 (2015).

40. Zhao, Y. *et al.* Spin-Layer Locking Induced Second-Order Nonlinear Effect in Centrosymmetric Crystals. https://www.researchsquare.com/article/rs-126567/v1 (2020).

41. He, X. W. *et al.* Anomalous Photogalvanic Effect of Circularly Polarized Light Incident on the Two-Dimensional Electron Gas in $Al_xGa_{1-x}N/GaN$ Heterostructures at Room Temperature. *Phys. Rev. Lett.* **101**, 147402 (2008).

42. Tang, C. G., Chen, Y. H., Liu, Y. & Wang, Z. G. Anomalous-circular photogalvanic effect in a GaAs/AlGaAs two-dimensional electron gas. *J. Phys. Condens. Matter* **21**, 375802 (2009).

43. Yu, J. *et al.* Photoinduced Inverse Spin Hall Effect of Surface States in the Topological Insulator $Bi_2Se_3$. *Nano Lett.* **17**, 7878–7885 (2017).

44. Mei, F. H. *et al.* Detection of spin-orbit coupling of surface electron layer via reciprocal spin Hall effect in InN films. *Appl. Phys. Lett.* **101**, 132404 (2012).

45. Zhang, Y. *et al.* Inverse spin Hall photocurrent in thin-film $MoTe_2$. *Appl. Phys. Lett.* **116**, 222103 (2020).

46. Zhang, Y. *et al.* Local Electric-Field-Induced Spin Photocurrent in $ReS_2$. *J. Phys. Chem. Lett.* **13**, 11689–11695 (2022).

47. Du, L. Comment on "Disentangling Orbital and Valley Hall Effects in Bilayers of Transition Metal Dichalcogenides". *Phys. Rev. Lett.* **127**, 149701 (2021).

48. Tao, Z. *et al.* Giant spin Hall effect in AB-stacked MoTe$_2$/WSe$_2$ bilayers. *Nat. Nanotechnol.* **19**, 28–33 (2024).

49. Roy, A., Guimarães, M. H. D. & Sławińska, J. Unconventional spin Hall effects in nonmagnetic solids. *Phys. Rev. Mater.* **6**, 045004 (2022).

50. Safeer, C. K. *et al.* Large Multidirectional Spin-to-Charge Conversion in Low-Symmetry Semimetal MoTe$_2$ at Room Temperature. *Nano Lett.* **19**, 8758–8766 (2019).

51. Cho, S. *et al.* Experimental Observation of Hidden Berry Curvature in Inversion-Symmetric Bulk 2H−WSe$_2$. *Phys. Rev. Lett.* **121**, 186401 (2018).

52. Kormányos, A., Zólyomi, V., Fal'ko, V. I. & Burkard, G. Tunable Berry curvature and valley and spin Hall effect in bilayer MoS$_2$. *Phys. Rev. B* **98**, 035408 (2018).

53. Le, C. & Sun, Y. Topology and symmetry of circular photogalvanic effect in the chiral multifold semimetals: a review. *J. Phys. Condens. Matter* **33**, 503003 (2021).

54. Dhara, S., Mele, E. J. & Agarwal, R. Voltage-tunable circular photogalvanic effect in silicon nanowires. *Science* **349**, 726–729 (2015).

55. Wang, B. M., Zhu, Y., Travaglini, H. C., Savrasov, S. Y. & Yu, D. Schottky Electric Field Induced Circular Photogalvanic Effect in Cd$_3$As$_2$ Nanobelts. Preprint at http://arxiv.org/abs/2210.03819 (2022).

56. Jung, E. *et al.* Unusually large exciton binding energy in multilayered 2H-MoTe$_2$. *Sci. Rep.* **12**, 4543 (2022).

57. Ruppert, C., Aslan, B. & Heinz, T. F. Optical Properties and Band Gap of Single- and Few-

Layer MoTe$_2$ Crystals. *Nano Lett.* **14**, 6231–6236 (2014).